\documentclass{PoS}
\pdfoutput=1
\usepackage{color,amsmath,amssymb,bbm,xifthen,tikz}

\title{Automated lattice perturbation theory and relativistic heavy quarks in the Columbia formulation}
\ShortTitle{Automated LPT and RHQ in the Columbia formulation}

\newcommand{\msbar}{{\overline{\rm MS}}}

\newcommand{\Qb}{{\overline{Q}}}

\author{\speaker{Christoph Lehner}\\
  RIKEN BNL Research Center, Brookhaven National Laboratory, Upton, NY 11973, USA\\
  E-mail: \email{clehner@quark.phy.bnl.gov}}

\abstract{We introduce a new computer algebra system optimized for use
  in lattice perturbation theory as well as continuum perturbation
  theory and a new framework to perform automated perturbative
  calculations on top of said computer algebra system. The new
  framework is used to tune the relativistic heavy quark action in the
  Columbia formulation at one loop in meanfield-improved perturbation
  theory.  Preliminary results for the matching and O$(a)$-improvement
  of heavy-light axial vector currents with light domain-wall quarks
  are also presented.}

\FullConference{The 30th International Symposium on Lattice Field Theory\\
                 June 24 -- 29,  2012\\
                 Cairns, Australia}

\begin{document}

\section{Introduction}
The automation of lattice perturbation theory (LPT) was pioneered by
L\"uscher and Weisz in Ref.~\cite{Luscher:1985wf}, where lists were
used to represent the information required to evaluate gluonic Feynman
diagrams numerically.  Using a similar data structure, these early
results were extended in Ref.~\cite{Hart:2004bd} to the case of
fermionic actions, enhanced in Ref.~\cite{Hart:2009nr} to also allow
for complex smearing, and further refined in
Refs.~\cite{Takeda:2008rr} and \cite{Hesse:2011zz} for various
fermionic actions.

Here we propose to use a more general data structure and represent the
gluonic and fermionic actions as well as operators and matrix elements
in a symbolic manner using a computer algebra system
(CAS).\footnote{The possibility of such an alternative approach was
  briefly mentioned already in Ref.~\cite{Luscher:1985wf}.}  This
approach has found wide use in the automation of continuum
perturbation theory, where many high-loop calculations are preformed
using FORM \cite{Vermaseren:2000nd}.  Unfortunately, FORM lacks
features to efficiently handle the complicated vertices that arise in
lattice actions.  In this work, we introduce a new CAS optimized for
LPT.  We present a flexible framework on top of said CAS that is
capable of performing perturbative calculations using a lattice
regulator as well as a continuum regulator and to generate
contractions for non-perturbative computations.

In the following we briefly describe the layout of the new framework
and present first results for the one-loop O$(a)$-improvement of
relativistic heavy quarks (RHQ) in the Columbia formulation
\cite{Christ:2006us} and the matching and O$(a)$-improvement of
heavy-light axial vector operators in the on-shell limit.

\section{Layout of the framework}
In this section we give a brief overview of the new framework that
consists of three C++ libraries: {\bf libcas}, a computer algebra
system, {\bf libqft}, a quantum field theory library that sits on top
of libcas, and {\bf libint}, a library for the numerical evaluation of
algebraic expressions of lattice loop integrals.

The library libcas parses algebraic expressions from text and XML
input, implements a pattern matching algorithm that is largely
compatible with FORM, and provides further methods to manipulate and
simplify general algebraic expressions.  One crucial feature of libcas
that allows for the efficient use in the context of LPT is the {\it
  function map}.  Compared to continuum perturbation theory the
vertices of lattice actions are complicated functions of momenta.  A
straightforward expansion of all terms that appear in a typical
Feynman diagram leads to an explosion of individual terms that is
prohibitive for LPT.  This issue is addressed within libcas by the
function map that allows to replace sub-expressions (such as vertices)
by references to unique functions in the function map in a
straightforward manner.  Hence the complicated momentum-dependence is
represented as a separate function that only has to be evaluated once,
and each instance of the vertex is represented by a mere function call
with respective momenta as arguments.  We make extensive use of the
function map not only to handle the complexity of lattice vertices but
also, e.g., to perform a general common subexpression elimination
within the lattice loop integrands.

The library libqft extracts vertices from text-representations of
lattice and continuum actions and operators, defines common lattice
actions, performs Wick contractions, and computes loop integrals in
dimensional regularization using Passarino-Veltman reduction
\cite{Passarino:1978jh}.

The library libint provides a convenient framework to numerically
evaluate the lattice loop integrands with a variety of integration
methods.

All three libraries are designed to be easily extensible and will
eventually be released as open source; a complete list of features
will be given in an upcoming publication \cite{Lehner:2012yz}.

\section{Example: Applications to heavy quark physics}
\begin{figure}[t]
  \centering
  \includegraphics[width=10.7cm]{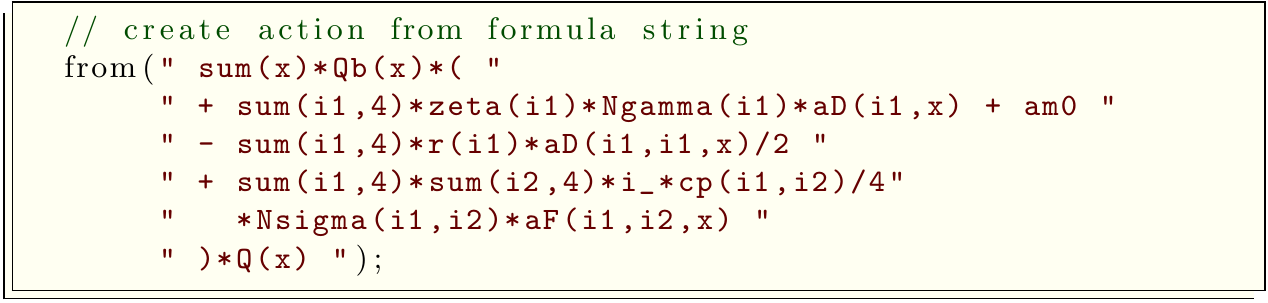}
  \caption{The definition of the RHQ action of
    Eq.~\protect\eqref{eqn:defrhq} within libqft is shown above.  The
    function {\it from} is defined within libqft and extracts the
    Feynman rules from the text representation of the action.
    Operators such as the covariant derivative {\it aD} or the
    field-strength tensor {\it aF} are defined within libqft.
    Special functions such as {\it sum} and symbols such as the
    complex {\it i\_} are defined within libcas.  The $\gamma$-algebra
    is represented by non-commuting functions {\it Ngamma} and {\it
      Nsigma}.  For a more detailed explanation of the above code, we
    refer to Ref.~\cite{Lehner:2012yz}.}
  \label{fig:libqftaction}
\end{figure}

The framework found its first applications within the heavy-quark
physics project of the RBC and UKQCD collaborations that uses RHQ in
the Columbia formulation to describe the heavy quarks.  Recent results
have been presented in Refs.~\cite{Kawanai:2012id} and
\cite{Witzel:2012pr}.  Relativistic heavy quarks, first proposed in
Ref.~\cite{ElKhadra:1996mp} and further refined in
Refs.~\cite{Aoki:2001ra} and \cite{Christ:2006us}, provide an
effective heavy-quark action for large quark masses that is smoothly
connected to a fully relativistic quark action as the quark mass
becomes small compared to the lattice cutoff.  The Columbia
formulation \cite{Christ:2006us} corresponds to the lattice action
\begin{align}\label{eqn:defrhq}
  S = \sum_{x} \Qb(x) \Biggl( ( \gamma_0 D_0- \frac12 D^2_0) +
  {\zeta}\sum\limits_{i=1}^3 ( \gamma_i D_i- \frac12 D^2_i)+ {m_0} +
  {c_P}\sum\limits_{\mu,\nu=0}^3 \frac{i}{4} \sigma_{\mu\nu}
  F_{\mu\nu} (x) \Biggr) Q(x)
\end{align}
with heavy-quark fields $Q$.  The parameters $m_0$, $\zeta$, and $c_P$
can be tuned to remove O$(a\vec{p})$ discretization errors in on-shell
quantities, where $a\vec{p}$ corresponds to the spatial momentum of
the heavy quark in lattice units.  Figure \ref{fig:libqftaction} shows
how this action is defined within libqft from a text representation.
If we allow for a field rotation
\begin{align}\label{eqn:deffieldrot}
  Q'(x) &= Q(x) + {d_1} \sum_{i=1,2,3} \gamma_i D_i Q(x)
\end{align}
with parameter $d_1$, we can match the quark fields $Q'$ to continuum
fields.  Such a field rotation, however, leaves the mass spectrum of
the theory invariant.  Therefore the parameters $m_0$, $\zeta$, and
$c_P$ can be tuned non-perturbatively without knowledge of $d_1$;
results for the tuning of bottom quarks were recently presented in
Ref.~\cite{Aoki:2012xaa}.

\begin{figure}[t]
  \centering
  \includegraphics[width=0.45\linewidth]{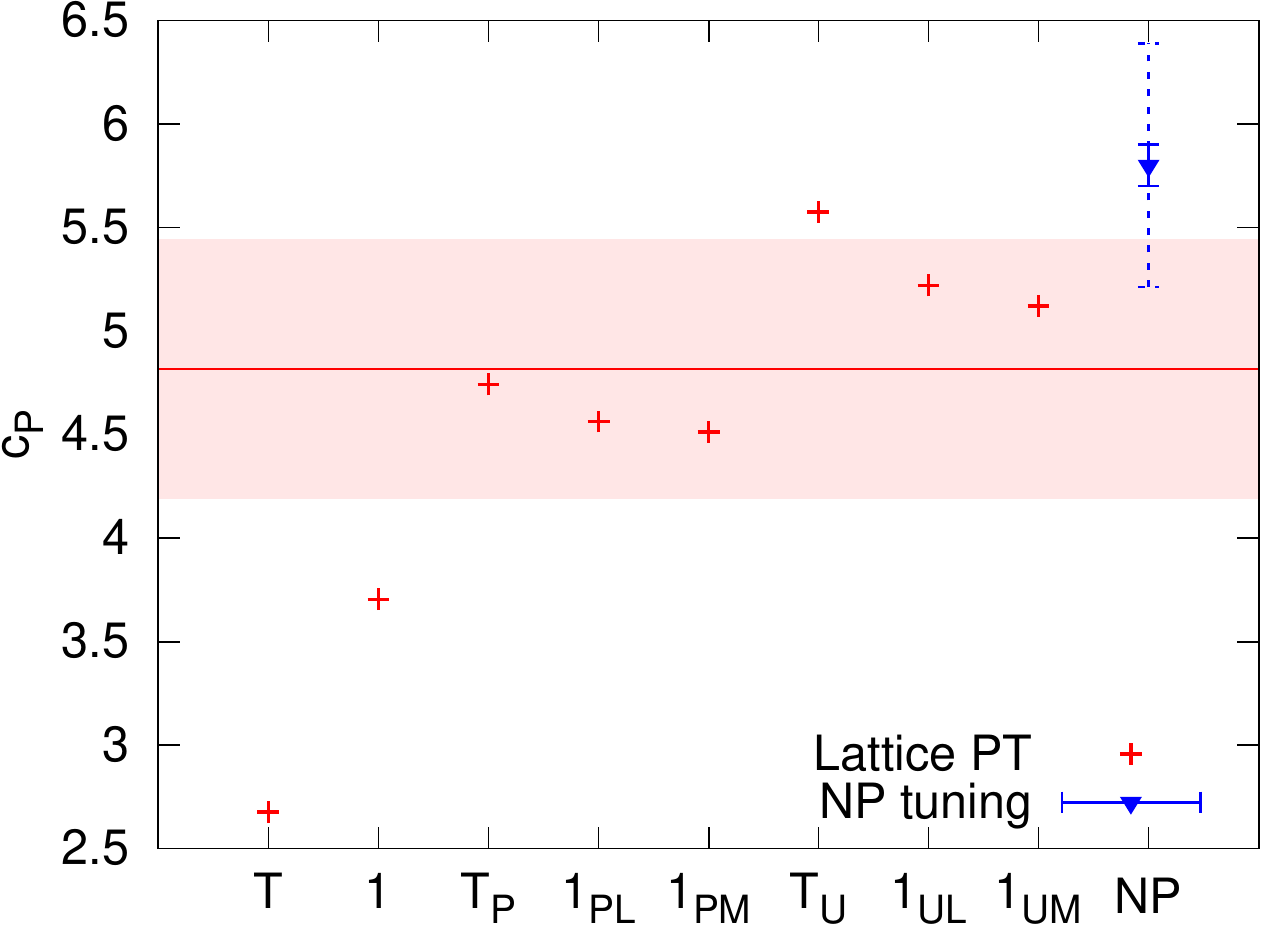}
  \hspace{0.05\linewidth}
  \includegraphics[width=0.45\linewidth]{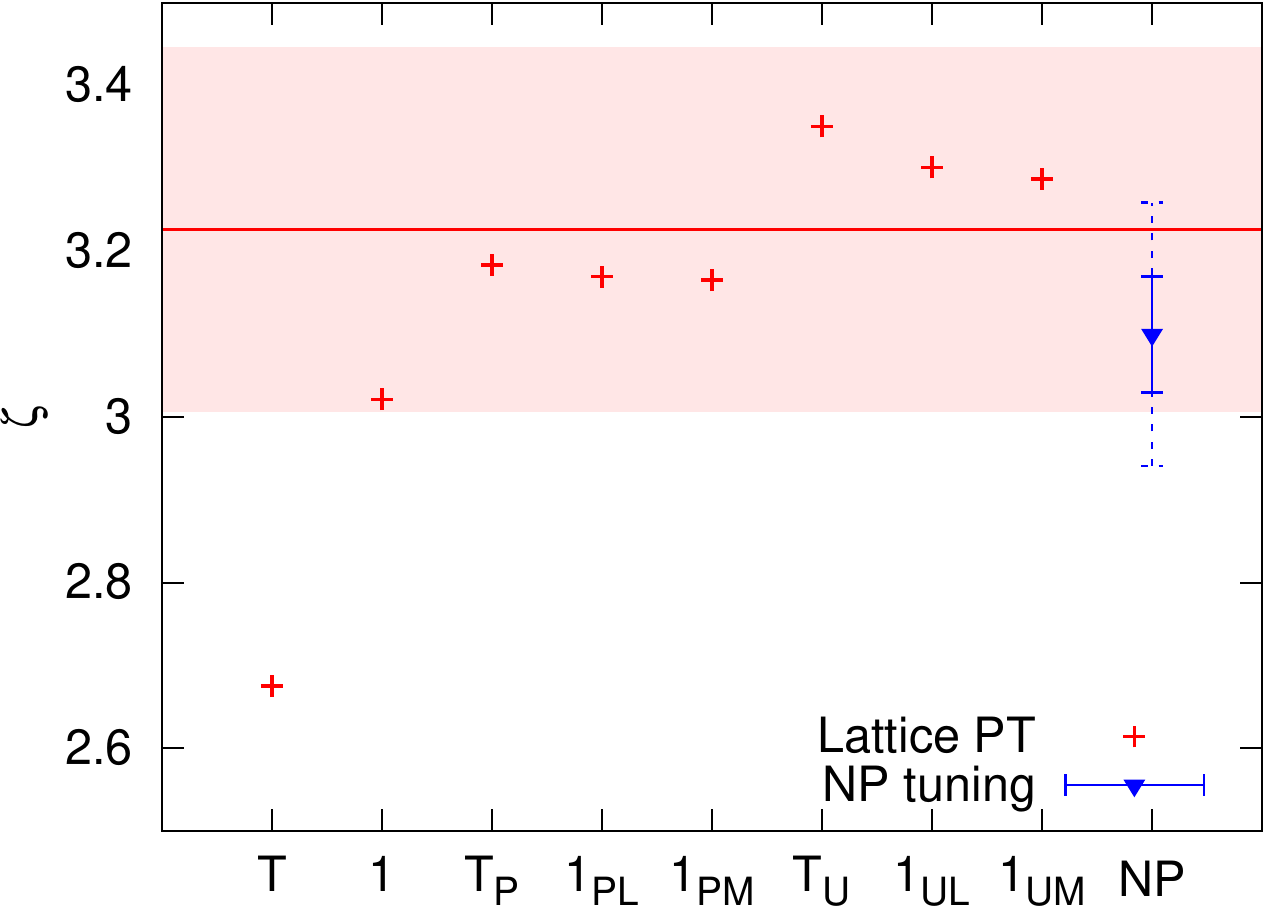}
  \caption{Results for the perturbative tuning of parameters $c_P$ and
    $\zeta$ on the $24^3$ ensembles of Ref.~\cite{Aoki:2012xaa} for
    bottom quarks.  We compare perturbative results with results of
    non-perturbative (NP) tuning.  Perturbative results are given at
    tree level (T) or at one-loop level (1).  The subscript P
    indicates that we use the average value of the plaquette for
    meanfield improvement, the subscript U indicates that the Landau
    gauge-fixed average link value was used for meanfield improvement.
    The subscript L denotes expansion in the bare lattice coupling,
    the subscript M denotes expansion in the $\msbar$ coupling
    constant at scale $1/a$, where $a$ is the lattice spacing.  The
    estimation of the errors is described in detail in
    Ref.~\cite{Aoki:2012xaa}.}
  \label{fig:tuning24}
\end{figure}

In order to test the framework, we also performed a perturbative
tuning of the parameters.  We match the bilinear
\begin{align}\label{eqn:defprop}
  S(p) &= \sum_q \langle Q'(p) \overline Q'(q)\rangle
\end{align}
to the continuum for on-shell momenta and determine $m_0$ from the
position of the pole, $\zeta$ from the dispersion relation, and $d_1$
from the spinor structure.  The parameter $c_P$ is obtained by
matching the three-point function
\begin{align}\label{eqn:defvert}
  \Lambda^a_\mu(p,q) &= \biggl[\sum_k \langle Q'(q) A^a_\mu(k)
  \overline Q'(-p) \rangle \biggr]_{\rm amp} = \biggl(\sum_{p'}\langle
  Q'(q) \overline Q'(p') \rangle\biggr)^{-1} \biggl(\sum_k\langle
  Q'(q) A^b_\nu(k) \overline Q'(-p)
  \rangle\biggr)\notag\\&\quad\times\biggl(\sum_{q'}\langle Q'(p)
  \overline Q'(q') \rangle\biggr)^{-1} [D(p+q)^{-1}]^{ba}_{\nu\mu}
\end{align}
to the continuum in the on-shell limit, where
\begin{align}
  D(k)^{ab}_{\mu\nu}= \sum_{k'}\langle A^a_\mu(k)A^b_\nu(k') \rangle
\end{align}
is the lattice gluon propagator with color indices $a$, $b$ and
Lorentz indices $\mu$, $\nu$.  Figure \ref{fig:tuning24} compares the
perturbative and non-perturbative results for $c_P$ and $\zeta$ on the
$24^3$ lattices of Ref.~\cite{Aoki:2012xaa}.  For more details of the
calculation, we refer to Refs.~\cite{Aoki:2012xaa} and
\cite{Lehner:2012yz}.  Within estimated errors, the perturbative and
non-perturbative results agree.

\begin{figure}[t]
  \centering
  \includegraphics[width=10.7cm]{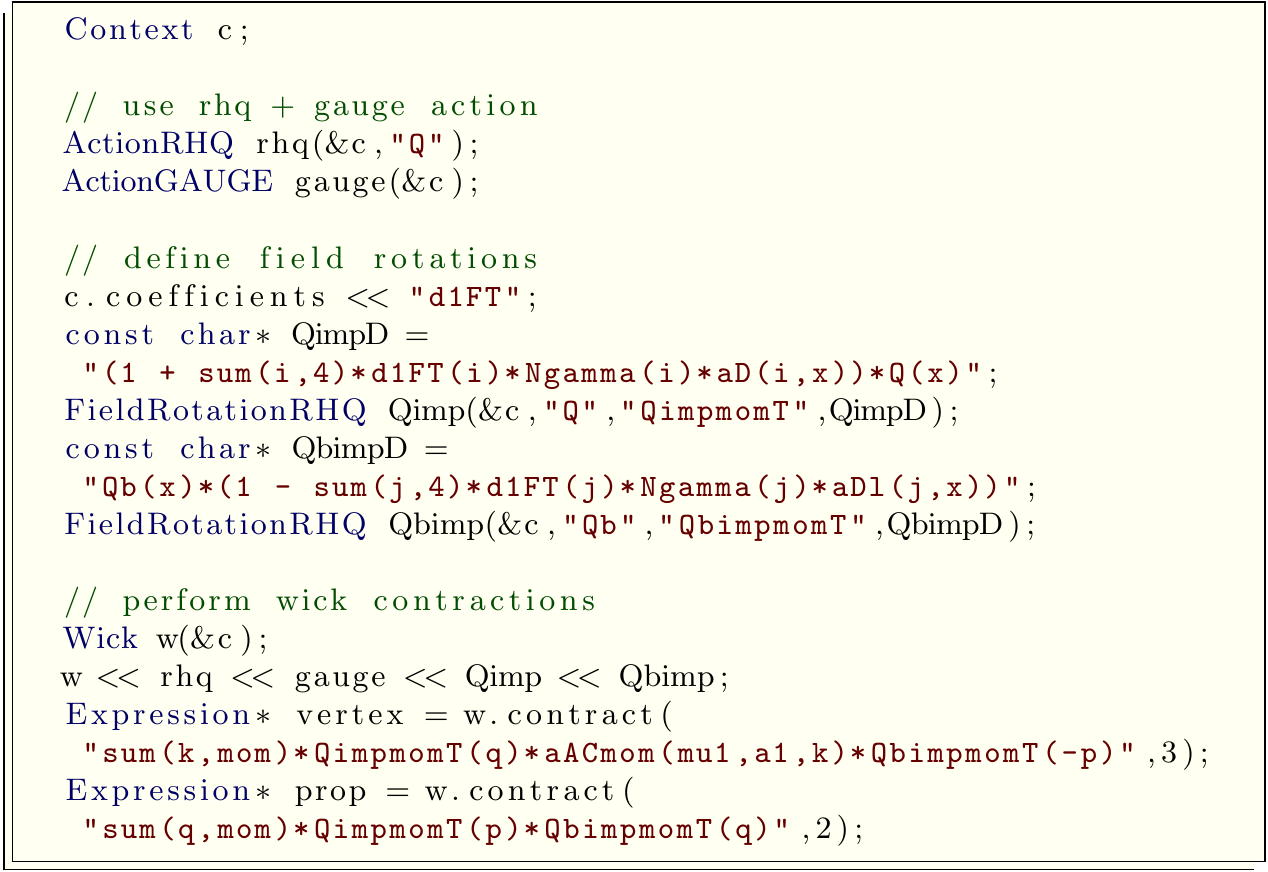}
  \caption{The results of Fig.~\protect\ref{fig:tuning24} are obtained
    using the code displayed above.  The code defines the field
    rotations of Eq.~\protect\eqref{eqn:deffieldrot} and generates the
    one-loop diagrams for the propagator of
    Eq.~\protect\eqref{eqn:defprop} and the vertex of
    Eq.~\protect\eqref{eqn:defvert}.  The resulting diagrams for the
    vertex are shown in Fig.~\protect\ref{fig:vertdiag}.  A more
    detailed explanation of the code shown above is given in
    Ref.~\protect\cite{Lehner:2012yz}.}
  \label{fig:libqftrhqmatch}
\end{figure}

\begin{figure}[t]
  \centering
  \includegraphics[width=12cm]{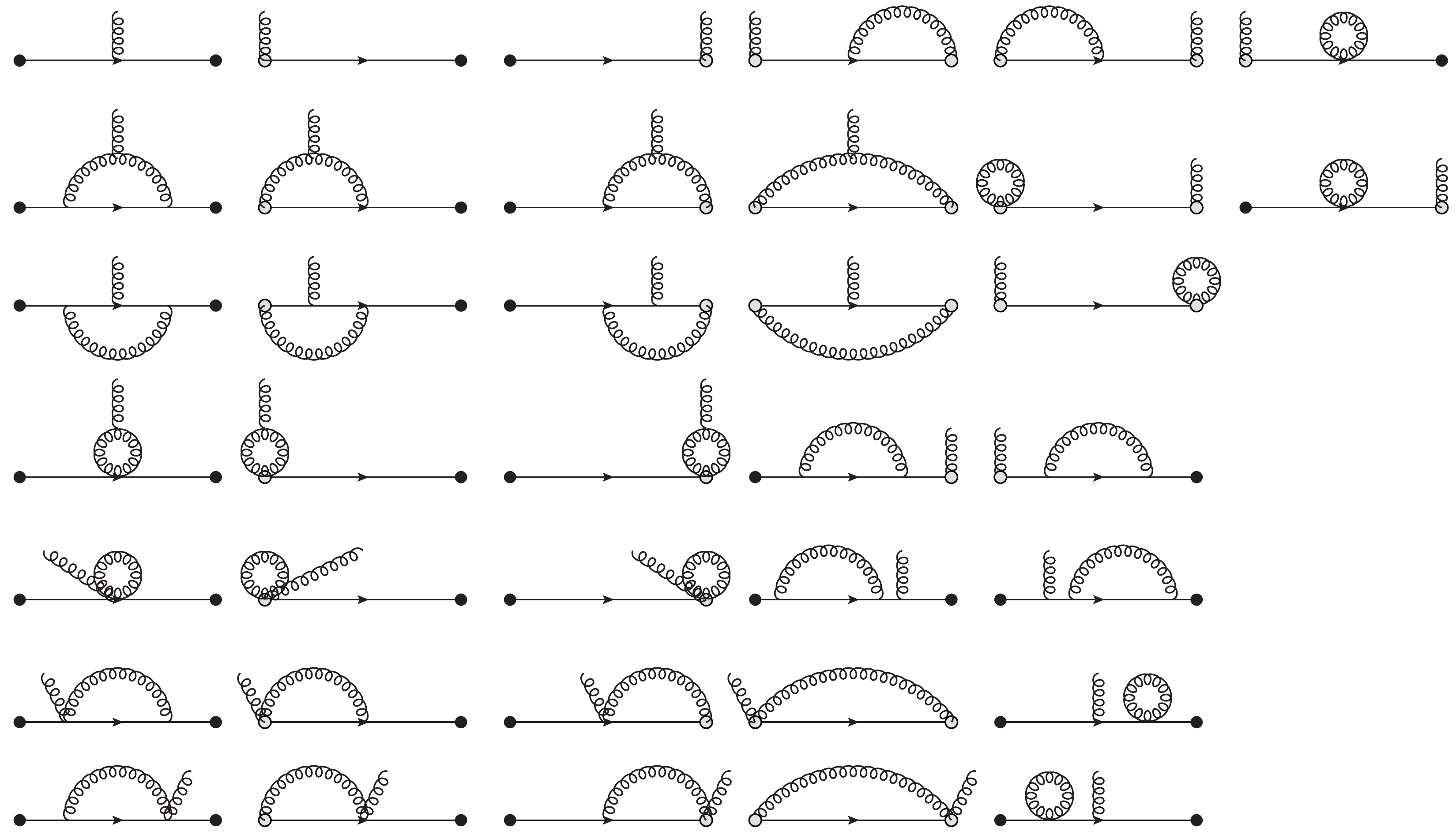}
  \caption{One-loop diagrams of the quark-gluon vertex of the RHQ
    action in the Columbia formulation defined in
    Eq.~\protect\eqref{eqn:defvert}.}
  \label{fig:vertdiag}
\end{figure}

The corresponding C++ code for the calculation of the necessary
diagrams is shown in Fig.~\ref{fig:libqftrhqmatch}, and the resulting
diagrams for the vertex are displayed in Fig.~\ref{fig:vertdiag}. The
one-loop integral for the propagator is evaluated to $10^{-3}$
relative accuracy on a regular desktop computer in less than ten
seconds.

\begin{figure}[t]
  \centering
  \includegraphics[width=10.7cm]{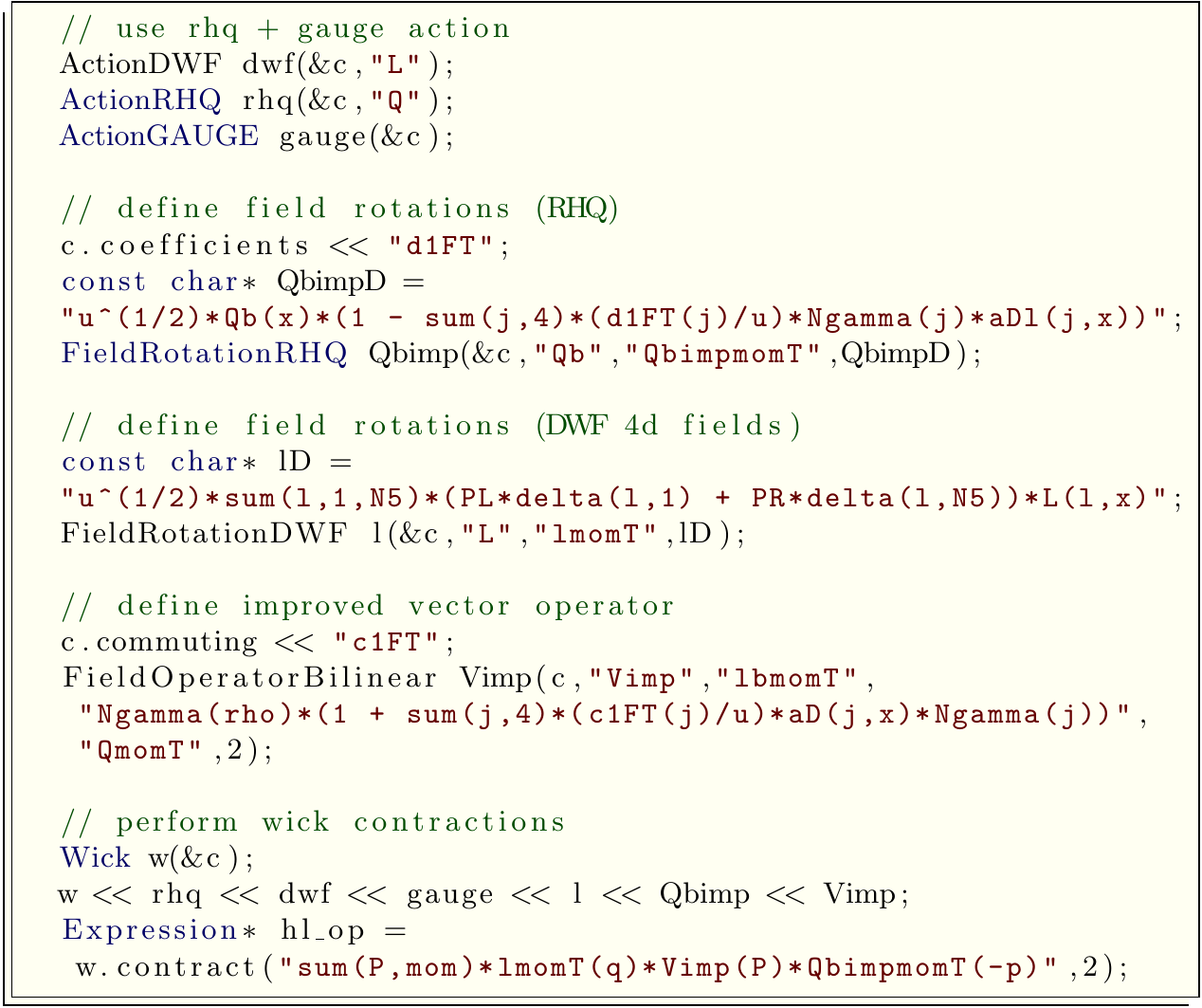}    
  \caption{The above code was used to obtain the results of
    Tab.~\protect\ref{tab:coef24}.  Field rotations are used to define
    the four-dimensional domain-wall quark.  For a more detailed
    explanation, we refer to Ref.~\cite{Lehner:2012yz}.}
  \label{fig:codehl}
\end{figure}

In Ref.~\cite{Witzel:2012pr} preliminary results for $f_B$ and
$f_{B_s}$ were presented with light domain-wall fermions.  The decay
constants are extracted from matrix elements of
\begin{align}
A_0^{\rm cont}(x) &= \rho_{bl} \sqrt{Z_V^{bb} Z_V^{ll}} \biggl[
\overline q(x) \gamma_0 \gamma_5 Q(x) +
c_1 \sum_{i=1,2,3}\overline q(x) \gamma_0 \gamma_5 \gamma_i D_i Q(x)\biggr]\,,
\end{align}
where $A_0^{\rm cont}$ is the temporal component of the heavy-light
axial vector operator in the continuum, $q$ is the domain-wall quark
field, and $Q$ is the relativistic heavy quark field.  The lattice
matrix element is related to the continuum version using the
non-perturbatively determined light-light vector matching factor
$Z_V^{ll}$ and the perturbatively determined coefficients $\eta_{bl} =
\rho_{bl} \sqrt{Z_V^{bb}}$ and $c_1$ with heavy-heavy vector matching
factor $Z_V^{bb}$.  The coefficient $c_1$ is tuned to remove
O($a\vec{p}$) errors.  In Tab.~\ref{tab:coef24} we present preliminary
meanfield-improved perturbative results at one loop for the
coefficients $\eta_{bl}$ and $c_1$ for the $24^3$ ensembles used in
Ref.~\cite{Witzel:2012pr}.  The perturbative results are based on the
code displayed in Fig.~\ref{fig:codehl}.

\begin{table}[t]
  \centering
  \begin{tabular}{ccccc}
    & $(\eta_{bl})^{(0)}$ & $(\eta_{bl})^{(1)}$ & $(c_1)^{(0)}$ & $(c_1)^{(1)}$ \\\hline
    Plaquette & $3.2615$ &   $-0.0099$ & $0.0727$ & $0.0257$ \\
    Landau link & $3.3195$ & $-0.0107$ & $0.0753$&  $0.0254$
  \end{tabular}
  \caption{Preliminary perturbative results for $\eta_{bl} = (\eta_{bl})^{(0)} + g^2 (\eta_{bl})^{(1)}$ and 
    $c_1 = (c_1)^{(0)} + g^2 (c_1)^{(1)}$, where $g$ is the strong coupling constant, for the $24^3$ ensemble used in Ref.~\cite{Witzel:2012pr}.
    The results are given for meanfield improvement using the average plaquette or the Landau gauge-fixed average link.}
  \label{tab:coef24}
\end{table}

\section{Concluding remarks}
A main virtue of the new framework presented in this paper is its
versatility through the use of the new CAS.  It provides a unified
environment to perform perturbative calculations with both lattice as
well as continuum regulators and is also able to generate contractions
for non-perturbative computations.  All expressions are stored in an
algebraic manner and can be modified in a straightforward way using
pattern-matching techniques.

We presented results of a first application in the context of the
heavy quark physics program of the RBC and UKQCD collaborations.  The
framework is designed with higher-loop calculations in mind; a
challenge that we plan to address soon.  A publication containing
further details is in progress \cite{Lehner:2012yz}.

\acknowledgments 

I would like to thank my colleagues of the RBC and UKQCD
collaborations for many useful discussions and acknowledge support by
the RIKEN FPR program.

\providecommand{\href}[2]{#2}\begingroup\raggedright\endgroup

\end{document}